\documentstyle[prl,aps,twocolumn,psfig]{revtex}
\begin{document}
\hyphenation{Ryd-berg}
\flushbottom
\draft
\title{Semiclassical interferences and catastrophes in the
ionization of Rydberg atoms by half-cycle pulses}
\author{G.~Alber$^1$ and O.~Zobay$^2$} 
\address{$^1$Abteilung f\"ur Quantenphysik, Universit\"at Ulm,
D--89069 Ulm, Germany\\
$^2$Optical Sciences Center, University of Arizona,
Tucson, Arizona 85721\\
\underline{to be printed in Phys. Rev. A
(rapid communication)}}
\maketitle
\begin{abstract}
A multi-dimensional
semiclassical description of excitation of a Rydberg electron
by half-cycle pulses is developed and applied to the study of energy-
and angle-resolved ionization spectra. Characteristic novel
phenomena observable
in these spectra such as interference
oscillations and semiclassical glory
and rainbow scattering are discussed and related to the 
underlying classical
dynamics of the Rydberg electron.
Modifications to the predictions of
the impulse approximation are examined
that arise due to finite pulse durations.
%The amplitude and phase information which is accessible
%in these angle-resolved ionization spectra might 
%offer new perspectives for probing highly excited quantum states
%of a Rydberg electron.
\end{abstract}
\narrowtext
High power, nearly unipolar electromagnetic field pulses
are a useful new spectroscopic tool in particular
for studying the dynamics of weakly bound Rydberg electrons.
%These half-cycle pulses (HCPs) give rise to interesting novel phenomena
%which cannot be observed with conventional laser pulses or microwave
%fields which involve many oscillatory cycles.
Various experimental \cite{JYB93,J96,RCSB96} and theoretical
\cite{RSB94,MC96} investigations on
the interaction between Rydberg atoms and HCPs
have already been performed.
In particular, the theoretical approaches which have been used
so far involve either fully quantum mechanical calculations, 
purely classical simulations,
or simplified one-dimensional model problems.
Fully quantum mechanical treatments are capable of yielding
quantitatively exact results but due to the large spatial extension of
highly excited Rydberg states they demand a huge computational effort.
Furthermore, they allow only for a restricted qualitative understanding
of the system, in general.
Purely classical approaches are of interest as they yield
insight into the underlying classical aspects of the motion of the
Rydberg electron. But they cannot deal properly with
quantum mechanical interference effects and are therefore also of
limited applicability. One-dimensional model problems, finally,
are not capable of describing all spatial aspects of the excitation
process. Thus, the natural question arises whether a proper three-dimensional
theoretical description of the excitation process can
be developed which is numerically
highly accurate even for very large principal quantum numbers,
which gives direct insight
into the underlying classical aspects of the dynamics of the Rydberg
electron and which is capable of dealing
with all quantum mechanical interference phenomena properly.

In this Communication it is shown that such a theoretical description
can be developed on the basis of a multidimensional semiclassical
description of the excitation process. This approach is easily
applicable to any spatial or temporal pulse form
of the exciting HCP and it is particularly accurate in the region of
high principal quantum numbers which
is difficult
to access by fully quantum mechanical calculations.
Based on this treatment novel oscillatory structures are
discussed which appear in energy- and angle-resolved ionization spectra
and which should be accessible to experimental observation
in view of the recently developed imaging method \cite{Helm}.
It is shown that these structures are caused by quantum mechanical
interferences between probability amplitudes that can be associated with
classical trajectories of the ionized Rydberg electron.
%A further generic
%feature is the appearance of
%semiclassical catastrophes of the glory and rainbow type.
In order to
demonstrate the numerical accuracy of the multidimensional
semiclassical approach a comparison with numerical results is presented
in the impulse approximation.
%Although, for the sake of simplicity, most of
%our subsequent discussion concentrates on the
%impulse approximation also
Modifications of these oscillatory structures due to finite
pulse durations are also discussed.
%As the presented multidimensional semiclassical description
%takes into account all the phases of relevant probability amplitudes
%it might also offer new perspectives
%for probing highly excited quantum states of 
%Rydberg electrons.

Let us consider a typical HCP-excitation process. Initially
an atom is prepared 
in a Rydberg state $|n_0l_0m_0 \rangle$ with
a large principal quantum number $n_0 \gg 1$ and a small
angular momentum quantum number $l_0 \ll n_0$.
Around $t=0$ a linearly polarized HCP with vector potential
$A({\bf x},t){\bf e}_z$ and duration $\tau$ interacts with the
Rydberg electron.
The time evolution of the state $\psi({\bf x}, t)$ of the Rydberg electron
is determined by the Schr\"odinger equation
\begin{equation}
i\dot{\psi}({\bf x}, t) = 
\{\textstyle{\frac 1 2}{[}-i\nabla_{\bf x}-  A({\bf x}, t){\bf e}_z{]}^2 -
V_C({\bf x})\}\psi({\bf x}, t) 
\label{S}
\end{equation}
with the effective electronic potential $V_C({\bf x})$
(Hartree atomic units with $e=\hbar=m_{\rm e}=1$ are used).
Assuming axial symmetry of $A({\bf x},t)$ around the $z$-axis the
$z$-component of the electronic angular momentum $L_z$ is conserved.
As long as
%the applied field strength is sufficiently weak,
%i.e.\ $| {\bf A}({\bf x}, t)|^2/2 \ll 1$, and
$n_0 \gg 1$
the solution of Eq.\ (\ref{S}) can be obtained
to a good degree of approximation with the help of semiclassical
methods. Thereby, one starts from the semiclassical expression for the
initial Rydberg state
\begin{eqnarray}
\langle {\bf x}_0 | n_0 l_0m_0\rangle &=&
A_{cl}({\bf x}_0) \{e^{i[S_0(r_0,\epsilon_0) - \pi/4]} +
{\rm c.c.}\}.
\label{semi}
\end{eqnarray}
The amplitude of this state is given by \cite{BS}
\begin{equation}
A_{cl}({\bf x}_0) = 
\frac{Y_{l_0}^{m_0}(\Theta_0,\phi_0)}{r_0 (n_0 - \alpha)^{3/2}
\sqrt{2\pi p(r_0,\epsilon_0)}}
\end{equation}
with the spherical harmonic $Y_{l_0}^{m_0}$,
the radial momentum
$p(r_0,\epsilon_0) = \sqrt{2(\epsilon_0 + 1/r_0)}$, 
and with the initial energy
$\epsilon_0 = -[2(n_0 - \alpha)^2]^{-1}$ ($\alpha$ denotes the quantum
defect of the Rydberg electron).
The phases of this state are determined by the classical eikonal
$S_0(r_0,\epsilon_0) = \int_0^r dr' p(r',\epsilon_0) - (l_0 + 1/2)\pi
+ \pi \alpha$.
Equation (\ref{semi}) is valid for distances
$r_0 = |{\bf x}_0|$
of the electron from the nucleus which are located outside the
core region and well inside the classically allowed region.
Semiclassically
one has to associate with this particular initial state a bi-valued
field of classical momenta
${\bf p}^{(\pm)}({\bf x}_0)=\pm \nabla_{{\bf x}_0}
S_0(r_0,\epsilon_0)$ which defines a Lagrangian manifold.
Generalizations to other types of initial states which lead to
other Lagrangian manifolds are possible according to the general
framework of multidimensional semiclassical approximations
\cite{Maslov}. Thus under the
influence of the HCP two classical trajectories ${\bf x}_{\pm}(t; {\bf
x}_0)$ emanate from each point ${\bf x}_0$. They are solutions
of the classical equations of motion of the Hamilton function
$H = [{\bf p} - {\bf A}({\bf x}, t)]^2/2 + V_C({\bf x})$ with
initial
conditions $[{\bf x}_0,{\bf p}^{(\pm)}({\bf x}_0)]$.
From Eq.\ (\ref{S}) and Maslov's generalization of the
Van-Vleck propagator
\cite{Maslov} one obtains
the probability amplitude of observing
the ionized Rydberg electron with final asymptotic
momentum ${\bf p}^{(f)}$, namely 
\begin{eqnarray}
\langle {\bf p}^{(f)}| \psi (t\to \infty)\rangle &=& \sum_{j}
P^{(cl)}_{j} e^{i[S_{j} + W_j({\bf x}_0^{(j)}) - \pi\mu_{j}/2 ]}.
\label{ampl}
\end{eqnarray}
As this result is understandable in a straightforward way
on physical terms the 
details of its derivation will be presented elsewhere
\cite{ZA98}.

The ionization amplitude of Eq.\ (\ref{ampl})
is expressed as a sum of contributions of all
classical trajectories $j$ of the Rydberg electron with initial
positions ${\bf x}_0^{(j)}$ which yield the asymptotic
momentum ${\bf p}^{(f)}={\bf p}_{\pm}(t\to \infty;{\bf x}_0, {\bf
p}^{(\pm)}({\bf x}_0))$.
The contribution of trajectory $j$ to the classical
angle- and energy-resolved
ionization amplitude is determined by
\begin{equation}
P_j^{(cl)} =\left| 
\frac{ dp_x^{(f)}\wedge dp_y^{(f)}\wedge dp_z^{(f)} }{
dx_0\wedge dy_0\wedge dz_0 }
\right|_{j}^{-1/2}
A_{cl}({\bf x}_0^{(j)}).
\end{equation}
The phases associated with these trajectories are determined 
by the Morse indices $\mu_{j}$, by the classical actions
\begin{equation}
S_{j}
= - \int_0^{\infty} dt\, 
{\bf x}_{j}(t)\cdot \frac{d{\bf p}^{(j)}}
{dt} - {\bf x}^{(j)}_0 \cdot {\bf p}_0^{(j)},
\end{equation}
and by the classical actions of the initial state
$W_j({\bf x}_0^{(j)}) = \pm[S_0(r_0^{(j)},\epsilon_0) - \pi/4]$.
For $W_j({\bf x}_0^{(j)})$ one has to choose the plus or the minus sign
depending on whether the initial radial momentum of the classical
trajectory $j$ is positive or negative.
The Morse index is equal to the number of 
sign changes of $
\frac{
dp_x^{(f)}\wedge dp_y^{(f)}\wedge dp_z^{(f)}
}{
dx_0\wedge dy_0\wedge dz_0
}\equiv {\rm Det}\frac{
\partial(p_x^{(f)}, p_y^{(f)}, p_z^{(f)})
}{\partial(x_0, y_0, z_0)}$
along the trajectory times their multiplicities \cite{Maslov}.
In terms of the probability amplitude of Eq.\ (\ref{ampl})
the semiclassical angle- and energy resolved ionization probability
is given by
\begin{eqnarray}
\frac{d^3P_{ion}}{d\epsilon \wedge d\Omega} &=& \sqrt{2\epsilon^{(f)}}
\left|\langle {\bf p}^{(f)}| \psi(t\to \infty)\rangle
\right|^2
\label{ion}
\end{eqnarray}
with $\epsilon^{(f)} = {\bf p}^{(f)\, 2}/2$ and $d\Omega \equiv \sin{\Theta_f}
d\Theta_f \wedge d\phi_f$. Neglecting all quantum mechanical interferences
between the contributing classical trajectories the classical
ionization probability
$d^3P^{(cl)}_{ion}/[d\epsilon \wedge d\Omega] = \sqrt{2\epsilon^{(f)}}
\sum_{j} (P_{j}^{(cl)})^2 $
is obtained. Without angular resolution this expression has already been
used successfully for explaining  particular aspects of ionization
by half-cycle pulses \cite{J96,RCSB96}.
Eqs.\ (\ref{ampl}) and (\ref{ion}) constitute a new, multidimensional
semiclassical framework within which 
all aspects of ionization by an HCP can be understood in terms of the
underlying classical dynamics of the ionized Rydberg electron.
They constitute a main result of this Communication.
It is worth noting that 
for all aspects which are sensitive to interferences
between different probability amplitudes, such as the angular
distributions discussed below, the inclusion of the proper
phases entering Eq.\ (\ref{ampl}) is crucial.
Furthermore, it should be mentioned that there exists a wealth
of alternative methods for solving semiclassical initial value problems
which have been pioneered by E. Heller and W. H. Miller \cite{review}.
However, so far these methods have been applied predominantly
to problems with explicitly time-independent Hamiltonians.
%A similar, simplified approach was
%used in a recent study of a one-dimensional HCP model system in Ref.\
%\cite{MC96}.

In order to demonstrate characteristic novel phenomena in the 
angle- and energy-resolved ionization spectra and in order
to show the accuracy of this semiclassical approach
let us consider first of all the case of an HCP with a duration short in
comparison to the classical orbit time of the Rydberg electron. In this
case the sudden ionization or impulse approximation
\cite{RSB94} may be invoked. Thereby, the influence of the
HCP on the Rydberg electron is described as a sudden momentum change of
magnitude $\Delta {\bf p} = -\int_{-\infty}^{\infty} dt\, {\bf E}(t)
=\Delta p\,{\bf e}_z$ at $t=0$ [${\bf E}(t)$ denotes the electric
field of the HCP at the position of the atom].
After the sudden excitation the
motion of the electron is governed by the
Coulomb potential $V_C({\bf x})$ alone, i.e.\ 
${\bf A}({\bf x}, t) \equiv 0$ for $t>0$ in Eq.\ (\ref{S}).
Thus, immediately after the HCP the initial momenta of
the classical trajectories appearing in Eq.\ (\ref{ampl})
are given by
${\bf p}^{(\pm)}({\bf x}_0) + \Delta {\bf p}$.
The positions ${\bf x}_0^{(j)}$
at which the electron is imparted a final energy $\epsilon^{(f)}$
are determined by energy conservation, i.e.
\begin{equation}
\epsilon_0+\Delta {\bf p}^2/2+{\bf p}^{(\pm)}({\bf x}_0^{(j)})\cdot \Delta
{\bf p}= \epsilon^{(f)}.
\label{energy}
\end{equation}
Furthermore, consistent with this approximation the classical actions
$S_{j}$ of Eq.\ (\ref{ampl}) have to be replaced by $S_{j} + \Delta {\bf p}
\cdot {\bf x}_0^{(j)}$.

As a general consequence of the semiclassical
description based on
Eqs.\ (\ref{ampl}) and (\ref{ion}) scaling properties of the underlying
classical dynamics of the Rydberg electron should manifest themselves
in the ionization spectra. On the classical level
the trajectories $({\bf x}(t),{\bf p}(t))$ and
$(\tilde{\bf x}(t),\tilde{\bf p}(t))$ resulting
from two spatially
homogeneous HCPs with field strengths ${\bf E}(t)$ and
$\tilde{{\bf E}}(t)=\gamma^{4}{\bf E}(\gamma^{3} t)$, $\gamma
>0$, are related by the scaling transformation
$\tilde{\bf x}(t) = \gamma^{-2}{\bf x}(\gamma^{3}t)$,
$\tilde{\bf p}(t) = \gamma {\bf p}(\gamma^{3}t)$.
%${\bf x} \to \gamma^{-2}{\bf x}, {\bf p} \to \gamma{\bf p}$
%which implies that
%$S_j \to \gamma^{-1} S_j$ for the associated classical
%actions.
In the impulse approximation 
this scaling relation has the far reaching consequence
that the character of the classical dynamics
depends on $\Delta {\bf p}^2$ and $\epsilon_0$ only through
the ratio
$[\Delta {\bf p}^2/(-2\epsilon_0)]$ so that
angle-resolved spectra
for $(\epsilon_0,\Delta p,\epsilon^{(f)})$ and $(\gamma^2 \epsilon_0,
\gamma \Delta p, \gamma^2 \epsilon^{(f)})$ 
will exhibit similar qualitative structure.
As an example 
%for the rich structure exhibited by
%energy- and angle-resolved ionization spectra
let us consider the case $\Delta {\bf
p}^2/(-2\epsilon_0)=6.25$ in more detail.
For the sake of comparison with fully quantum mechanical results
the corresponding examples are calculated for $n_0=50$, i.e.\ $\Delta
p=0.05$ a.u., $l_0=0$, and $\alpha=0$.
From Eq.\ (\ref{energy})
two characteristic dynamical regimes can be distinguished according to
whether $\epsilon^{(f)}$ is smaller or larger than the critical
energy $\epsilon_0 + \Delta
{\bf p}^2/2$ $(=28.6$ meV in this example). % \cite{comment}.
In both cases the dynamics of the ionization process are very different.

\centerline{\psfig{figure=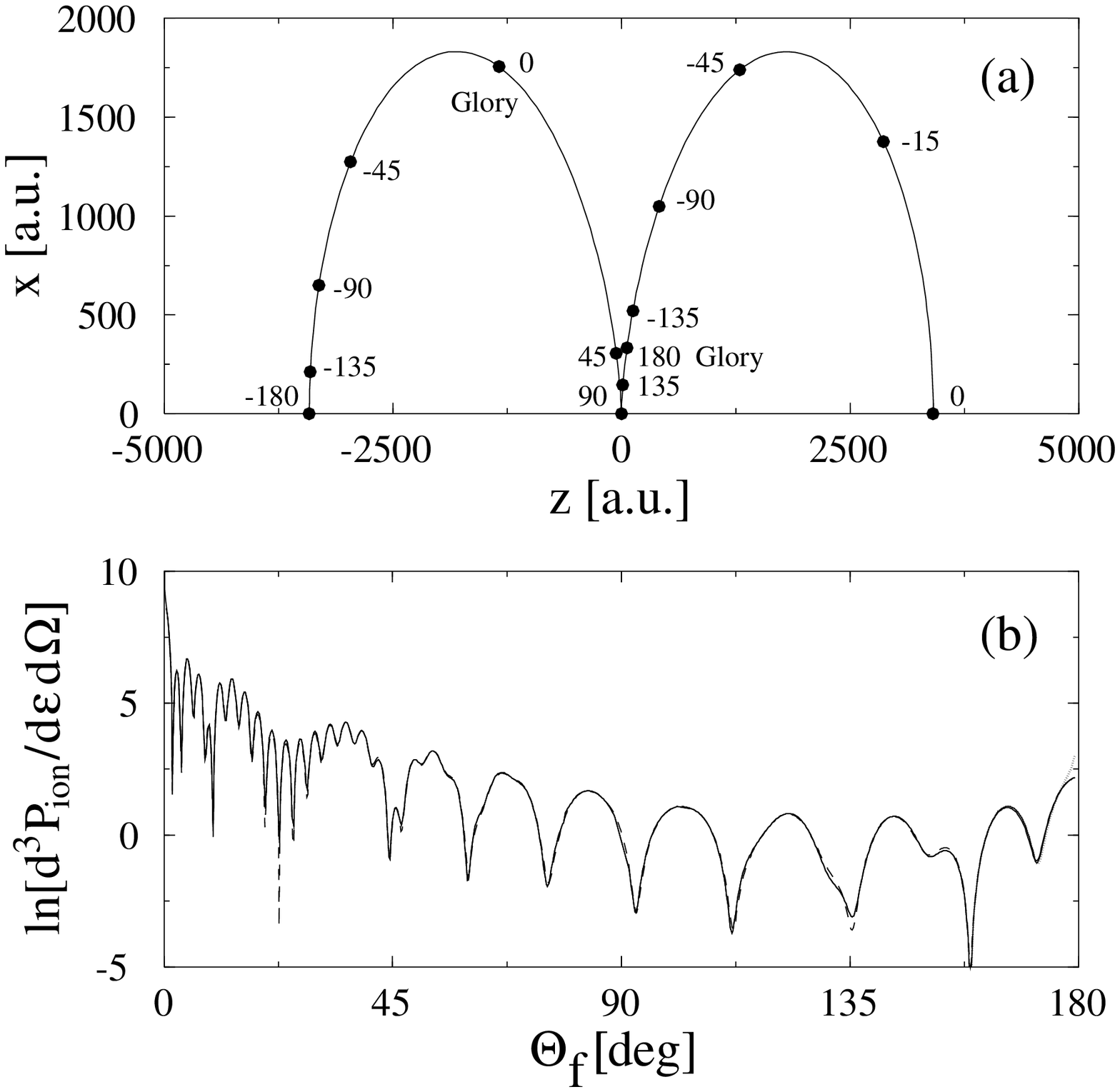,width=8.6cm,clip=}}
\begin{figure}
\caption{
(a) Full curve: initial positions of classical trajectories with
$n_0=50$, $\Delta p=0.05$ a.u., and $\epsilon^{(f)}=10$ meV according to
Eq.\ (\protect\ref{energy}). The figure has to be continued into three
dimensions by rotation around the $z$-axis. The numbers indicate final emission
angles for specific initial positions. Angle counted counterclockwise
from the $z$-axis.
(b) Angular distribution of the ionized electron
$\ln\{d^3P_{ion}/(d\epsilon d\Omega) [{\rm a.u.}] \}$ in the
sudden ionization approximation for the parameter values given above:
quantum mechanical result (full), uniform semiclassical (dashed)
and primitive semiclassical (dotted).}
\label{Fig1}
\end{figure}
Classical trajectories with $0 < \epsilon^{(f)} < \epsilon_0 + \Delta
{\bf p}^2/2$ start with negative (positive) radial momenta from the
positive (negative) $z$-half plane. Typical initial positions and final
angles of trajectories which according to Eq.\ (\ref{energy}) yield a
particular final energy $\epsilon^{(f)}$ are depicted in Fig.\
\ref{Fig1}(a) for a case with $\epsilon^{(f)}=10$ meV. Close inspection
of Fig.\ \ref{Fig1}(a) shows that for each final emission angle
$\Theta_{f}$ 
there are three different corresponding initial positions.
Therefore, one expects that the energy- and
angle-resolved ionization probability exhibits a characteristic
quantum mechanical interference structure which originates from these
three contributing trajectories. Furthermore, one notices that
there are classical trajectories with $\Theta_f = 0^{\circ}
(180^{\circ})$ whose initial positions do not lie on the $z$-axis.
The existence of these trajectories and the axial symmetry of
the dynamics around the polarization axis of the HCP give rise to the
semiclassical glory phenomenon \cite{Berry}. It manifests itself in a
divergence of the classical ionization
density $P^{(cl)}$ at $\Theta_f =
0^{\circ}$ and $\Theta_f =180^{\circ}$
which originates from the coalescence of 
two axially symmeytric families of classical trajectories. 
However, the resulting divergency of the primitive semiclassical
result of Eq.(\ref{ampl}) can be removed
with the help of semiclassical uniformization
methods \cite{ZA98,Berry}.
\centerline{\psfig{figure=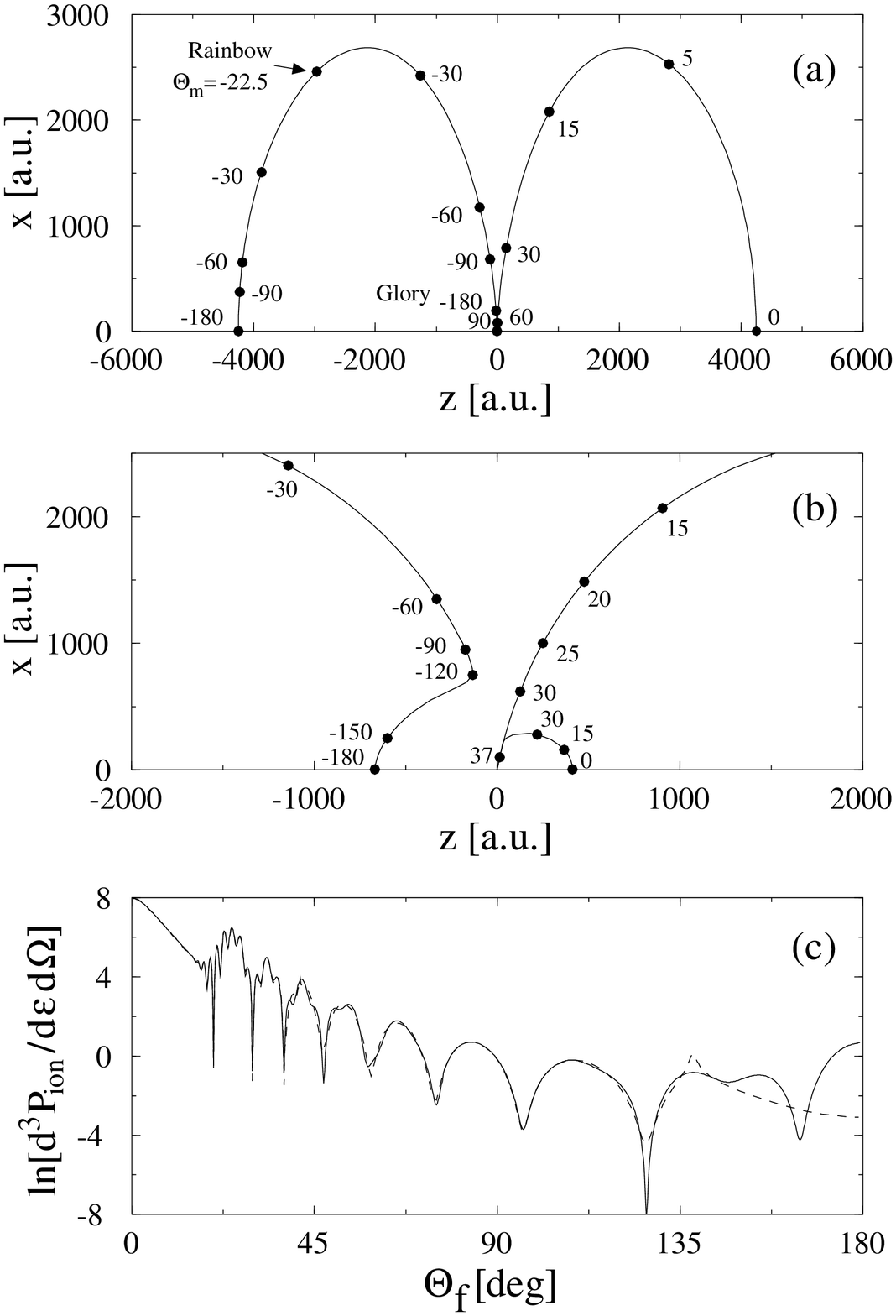,width=8.6cm,clip=}}
\begin{figure}
\caption{(a) Same as Fig.\ \protect\ref{Fig1}(a) for $\epsilon^{(f)}=40$
meV. (b) Same as (a) for the pulse of finite duration described in the
text. Shown is the region with the most significant modifications.
Outside this region the figure would be similar to (a).
(c) Angular distributions for the cases of Figs.\
\protect\ref{Fig2}(a) (full curve) and (b) (dashed).}
\label{Fig2}
\end{figure}
In Fig.\ \ref{Fig1}(b) the resulting energy- and angle-resolved
ionization spectrum is depicted.
In the sudden ionization approximation
the result of a full quantum mechanical calculation (full curve) is
compared with the corresponding semiclassical uniform result
(dashed curve)
and the
primitive semiclassical result based on Eq.(\ref{ampl}) (dotted curve).
The close agreement between the semiclassical and exact results
is apparent. 
The quantum interference between the contributing
classical trajectories  leads to characteristic oscillations of the
angular distribution. The angular distribution is concentrated
around the glory angles
$\Theta_f = 0^{\circ},\, 180^{\circ}$ where the
primitive semiclassical result diverges. 

The characteristic behavior of classical trajectories with
energies $1 \gg \epsilon^{(f)} > \epsilon_0 + \Delta {\bf p}^2/2$
is depicted in Fig.\ \ref{Fig2}(a).
In this case trajectories originating
from the positive (negative) $z$-half plane start with positive
(negative) radial momenta. From Fig.\ \ref{Fig2}(a) one notices 
an extremum of $\Theta_f$ on the curve of constant $\epsilon^{(f)}$. At
the extremum final angle $\Theta_m$ the contributions of two classical
trajectories
coalesce and give rise to a semiclassical rainbow phenomenon and a
divergence of the primitive semiclassical approximation.
Again this 
divergence can be removed by uniformization methods \cite{ZA98,Berry}.
In addition to this rainbow phenomenon glory scattering appears at
$\Theta_f =180^{\circ}$. However, the glory phenomenon at
$\Theta_f =0^{\circ}$ which is present in the case of low electron energies 
has now disappeared. The full curve of Fig.\ \ref{Fig2}(c) depicts
the uniform semiclassical result for $d^3P_{ion}/[d\epsilon \wedge d\Omega]$
in the sudden ionization approximation. It is again almost
indistinguishable from the corresponding quantum mechanical calculation
which is not shown. For small $\Theta_f$ the
spectrum is smooth as there is only one contributing trajectory. Around
$\Theta_m$ interference oscillations set in which reflect the
appearance of two further trajectory classes. The three
contributing trajectories persist up to
$\Theta_f =180^{\circ}$ where glory
scattering is observed in the impulse approximation.

Finally, let us discuss the influence of realistic
pulse forms on these ionization spectra with the help of
Eqs.\ (\ref{ampl})
and (\ref{ion}). Depending on the pulse duration and the spatial pulse
profile various modifications of these angular distributions are 
expected to appear. They will be studied systematically in 
forthcoming work \cite{ZA98}. As an example, let us concentrate here on
effects of finite pulse durations.
Figure \ref{Fig2}(b) depicts the modifications
in the distribution of initial
positions and final angles due to a moderately long HCP with pulse
duration $\tau=0.01T_{cl}$ ($T_{cl}=2\pi n_0^3$ is the
classical Kepler period of the Rydberg electron)
and pulse form ${\bf E}(t)={\bf e}_z E_0{\rm exp}[-(t/\tau)^8]$. The
amplitude $E_0$ is adjusted so that $|\Delta {\bf p}| =0.05{\rm\ a.u.}$
as in the previous cases. It turns out that trajectories starting
close to the nucleus are most sensitive to the pulse shape of an HCP.
Three major modifications in the energy-and angle resolved spectra
can be recognized: (i) trajectories with initially outgoing 
radial momenta lead to final angles with $|\Theta_f| < 37.5^{\circ}$,
only. However, this hardly affects the spectrum [dashed curve in Fig.\
\ref{Fig2}(c)] as the classical weight of this trajectory
class decreases
rapidly with increasing $\Theta_f$ already in the impulse 
approximation. (ii) A new class of trajectories with initially incoming
radial momenta appears in the positive $z$-half plane. They have final
angles $|\Theta_f| < 42^{\circ}$ and give rise to small oscillations
in the spectrum at small angles. (iii) The
most profound changes are due to trajectories in the negative
$z$-half plane. Their initial positions change significantly so that
the glory effect
at $\Theta_f = 180^{\circ}$ disappears. Thus, the spectrum
is most sensitive
to pulse shape effects at large emission angles. The rainbow
effect at small emission angles
is rather insensitive to the pulse shape.

In summary, a new multidimensional
semiclassical treatment of excitation
of a Rydberg electron by a HCP has been presented.
Based on this approach
it has been shown that energy-resolved angular distributions
of the ionized Rydberg electron exhibit characteristic
oscillations which can be interpreted naturally as contributions of
interfering classical trajectories.
These novel quantum mechanical interference features are
dominated by semiclassical catastrophes of the glory and rainbow
type.
As all quantum mechanical phases are taken into account properly
this semiclassical approach might also become a useful theoretical tool 
in problems concerning the reconstruction of quantum states of
Rydberg electrons with the help of HCPs.

Support by the Deutsche Forschungsgemeinschaft
%within the SPP
%``Zeitabh\"angige Ph\"anomene"
, by the
U.S.\ Office of Naval Research Contract No.\ 14-91-J1205, and by the
U.S.\ Army Research Office is acknowledged.
G. A. thanks H.~P.~Helm and P.~U.~Jepsen
for stimulating discussions.

\end{document}